\documentclass[11pt]{article}
\usepackage{moriond,epsfig}

\def\Journal#1#2#3#4{{#1} {\bf #2}, #3 (#4)}

\def\NIMA{{\em Nucl. Instrum. Methods} A}

\def\PLB{{\em Phys. Lett.}  B}
\def\PRL{\em Phys. Rev. Lett.}
\def\PRD{{\em Phys. Rev.} D}

\def\be{\begin{equation}}
\def\ee{\end{equation}}
\def\bea{\begin{eqnarray}}
\def\eea{\end{eqnarray}}

\begin{document}
\vspace*{4cm}
\title{MEASUREMENTS OF TOP QUARK PROPERTIES}

\author{ L. CERRITO (for the CDF and D0 Collaborations)}

\address{Department of Physics, Queen Mary University of London \\ Mile End Road,
London E1 4NS, England }

\maketitle\abstracts{Preliminary results on the measurement of four selected properties of the top quark are presented. The relative fraction of $t\bar{t}$ production through gluon fusion has been measured in the $t\bar{t}$ dilepton decay channel by the CDF Collaboration as $F_{gg}=0.53^{+0.36}_{-0.38}$. Using an integrated luminosity of 2.7 fb$^{-1}$ collected with the CDF II detector, we also determine the $t\bar{t}$ differential cross section with respect to values  up to $\sim$1 TeV of the $t\bar{t}$ invariant mass. We present a model-independent measurement of the helicity of $W$ bosons produced in top quark decays, using an integrated luminosity of up to 2.7 fb$^{-1}$ collected by the D0 detector, and find the fraction of longitudinal $W$ bosons $f_0=0.49\pm0.14$, and the fraction of right-handed $W$ bosons $f_+=0.11\pm0.08$. Finally, we measure the parton level forward-backward asymmetry of pair produced top quarks using an integrated luminosity of 3.2 fb$^{-1}$ collected with the CDF II detector, and find $A_{FB}=0.19\pm0.07$. All results are consistent with the predictions of the standard model.}

\section{Introduction}
The existence of the top quark has been confirmed in 1995 at the Tevatron collider \footnote{The Tevatron is a proton-antiproton synchrotron accelerator colliding beams at a center-of-mass energy of 1.96 TeV (Run II) in two locations (CDF and D0). Each experiment has to date recorded an integrated luminosity of $\sim$~6 fb$^{-1}$. The Tevatron operated until 1998 (Run I) at a center-of-mass energy of 1.8 TeV.}, and the top established within the standard model (SM) as the weak isospin partner of the bottom quark, with spin 1/2, interacting via all known forces, characterized by values of mass and width of $\sim 175$ GeV/$c^2$ and $\sim$1.5~GeV/$c^2$ respectively and decaying almost exclusively to $Wb$. Since the beginning of the Run II phase at the Tevatron in 2001, data samples of increasing numbers of top quark decays have been used to measure the top quark mass~\cite{topmass} ($m_t$) and the pair production cross section~\cite{topxsec} ($\sigma_{t\bar{t}}$) with  progressively greater precision. There is however a wide class of properties, such as the top quark production mechanism, the forward-backward asymmetry and the differential cross-section for example, whose experimental determination, limited by the statistical uncertainties, is only recently approaching the accuracy needed to test the SM predictions. A comprehensive study of top quark production and decay properties is crucial in view of the peculiar large value of $m_t$ and the special role the top may play in extensions of the SM. This article presents an overview of those properties  of top that have been determined experimentally, indicating the relevance they have in and beyond the SM, and the precision reached by the latest measurements.

\section{\boldmath Properties of Top Quark Pair Production} 
\noindent {\it \underline {$gg\rightarrow t\bar{t}$ Fraction}}\\
According to the SM, the production of $t\bar{t}$  at the Tevatron proceeds via $q\bar{q}$ annihilation ($q\bar{q}\rightarrow t\bar{t}$) or $gg$ fusion ($gg\rightarrow t\bar{t}$), the latter accounting for $\sim (15\pm5)$\% of the total with the uncertainty primarily due to the parton density functions (PDFs~\cite{gg1,gg2}). As a consequence of the short top quark lifetime~\cite{miao} ($\sim 5\times 10^{-24}$ s), the spin and kinematic information of the top are preserved in its decay products so that the two production mechanisms lead to different kinematic properties of the final-state particles. In general, top quark pairs produced via $q\bar{q}$ annihilation ($gg$ fusion) tend to be more central (forward) in the detector, with like (un-like) spins. Disagreement with the SM prediction could reveal the existence of new mechanisms of top-quark production or decay, such as a new vector particle associated with top color~\cite{topcolor} or a $t\rightarrow H^+b$ signature~\cite{higgs+}.
We report a measurement based on the azimuthal correlation of charged leptons in the $t\bar{t}$ dilepton decay channel. The signature of $t\bar{t}$ dilepton events consists of two high transverse momentum leptons from $W$ decay, large missing transverse energy due to two missing neutrinos, and two jets originating from $b$-quarks. The analysis uses an integrated luminosity of 2.0~fb$^{-1}$ collected with the CDF II~\cite{cdf} detector, selecting a total of 145  candidate $t\bar{t}$ events with 49.4$\pm$7.8 expected from background sources~\cite{ggeventsel}. The distribution of the azimuthal opening angle, $\Delta\phi$, of the charged leptons is determined separately for the $gg$, $q\bar{q}$ initiated processes and for each source of background events.
These templates are fit to the data distribution of $\Delta\phi$ using an unbinned likelihood fit, with the $gg$ fraction ($F_{gg}$) set as a free fit parameter. The fit returns:
$F_{gg} = 0.53^{+0.35}_{-0.37} ({\rm stat.})^{+0.07}_{-0.08}({\rm syst.}),$
with the statistical contribution dominating the uncertainty, and with the largest systematic uncertainty due to the background modeling. While this measurement is the first determination of $F_{gg}$ using the dilepton decay channel, the most precise measurement to date~\cite{ggf} is $F_{gg}=0.07^{+0.15}_{-0.07}$, obtained with the much larger $\ell +$ jets sample of top quarks, yielding an accuracy comparable to the expected magnitude of the $gg$ fraction.

\noindent {\it \underline {$d\sigma/dM_{t\bar{t}}$ Differential Cross Section}}\\
The study of the invariant mass of the top and anti-top system ($M_{t\bar{t}}$) in $p\bar{p}$ collisions is of relevance because $M_{t\bar{t}}$ is sensitive to a very broad class of models beyond the SM. These may produce distortions to the $d\sigma_{t\bar{t}}/M_{t\bar{t}}$ production spectrum ranging from narrow resonances to broad interferences~\cite{mtt_models}. The first measurement~\cite{mtt} of $d\sigma_{t\bar{t}}/M_{t\bar{t}}$ up to values of $M_{t\bar{t}}\simeq 1$~ TeV, using 2.7 fb$^{-1}$ of CDF II data, indicates good agreement with the SM expectation.

\noindent {\it \underline {Forward-Backward Asymmetry in $t\bar{t}$ Production}}\\
In Quantum Chromo Dynamics, a small charge asymmetry $A_C$ arises in next-to-leading order (NLO) $t\bar{t}X$ production. Because the strong interaction is invariant under charge conjugation, $A_C$ is equivalent to a forward-backward asymmetry $A_{FB}$ and in the Tevatron $p\bar{p}$ rest frame is calculated~\cite{afbth1} to be $A_{FB}=(5.0\pm1.5)$\%, where the uncertainty is driven by the size of the corrections at higher orders. Measurements of $A_{FB}$ have previously been reported~\cite{afbd0,afbcdf}. We present a new measurement of the parton level $A_{FB}$ of pair produced top quarks updating the latter result and increasing the dataset analyzed therein from 1.9~fb$^{-1}$ to 3.2~fb$^{-1}$ of $p\bar{p}$ collision data. A sample of 776  candidate $t\bar{t}$ events is selected in the $\ell +$ jets decay channel, where one top decays semi-leptonically ($t\rightarrow \ell\nu b$) and the other hadronically ($t\rightarrow q\bar{q}b$). We expect a background of 167$\pm$33 events.
We study the rapidity $y_{\rm had}$ of the hadronically-decaying top (or anti-top) system, tagging the charge with the lepton sign $Q_\ell$ from the leptonically decaying system. The hadronically decaying system is chosen because its direction is better reconstructed than the leptonic one, which is complicated by the missing energy of the neutrino.  In the $p\bar{p}$ center-of-mass frame we measure the asymmetry in $-Q\cdot y_{\rm had}$:
\begin{equation}
A_{FB}^{p\bar{p}} = \frac{N\left(-Q\cdot y_{\rm had}>0\right)-N\left(-Q\cdot y_{\rm had}<0\right)}{N\left(-Q\cdot y_{\rm had}>0\right)+N\left(-Q\cdot y_{\rm had}<0\right)},
\end{equation}
where $N$ indicates number of events. We subtract background and perform a model-independent correction for acceptance and reconstruction dilutions in order to find the asymmetry at the parton level. We find $A_{FB}^{p\bar{p}}=0.193\pm0.065~({\rm stat.})\pm0.024({\rm syst.})$, consistent with previous results and higher than, but in agreement with, the asymmetry expected in QCD at NLO.

\section{\boldmath Top Quark Decays and Other Properties}
\noindent {\it \underline {Helicity of W bosons in $t\bar{t}$ Decays}}\\
Charged current weak interactions proceed via the exchange of a $W^{\pm}$ boson and are described by a vertex factor that has a pure vector minus axial vector $\left( V-A\right)$ structure~\cite{v-a}. As a consequence, the $W$ bosons from the top-quark decay ($t\rightarrow bW^+$) are dominantly either longitudinally polarized or left-handed, while right-handed $W$ bosons are heavily suppressed ($O(10^{-4})$). At leading order (LO) in perturbation theory, the fraction of longitudinally polarized $W$ bosons is given by~\cite{helic} $f_0 = \frac{\Gamma(W_0)}{\Gamma(W_0)+\Gamma(W_-)+\Gamma(W_+)}\approx\frac{m_t^2}{2m_W^2+m_t^2}$, where $W_0$ and $W_\pm$ indicate longitudinally and transversely polarized $W$ bosons, respectively. Assuming a top-quark mass of 175 GeV/$c^2$, the SM predicts $f_0\simeq$ 0.7, while the fraction of left-handed $W$ bosons is $f_-\simeq 0.3$. A significant deviation, at level exceeding 1\% from the predicted value for $f_0$ or $f_+$ could indicate new physics, such as a possible $V + A$ component in the weak interaction or other anomalous couplings at the $Wtb$ vertex. 
The polarization of the $W$ can be inferred from the kinematics of the final state particles. In particular, we exploit the distribution of the angle ($\theta^*$) between the down-type fermion and top quark momenta in the $W$ boson rest frame. This measurement uses a data sample recorded with the D0 experiment~\cite{D0} at the Tevatron; the D0 detector underwent significant enhancement in 2006 so that we split the data into ``Run IIa'' and ``Run IIb'' subsamples, denoting data recorded before and after the detector enhancement. The data analysis of the Run~IIb subset is combined here with the $\sim$1 fb$^{-1}$ Run IIa measurement~\cite{D0helic} for a total integrated luminosity of 2.2--2.7 fb$^{-1}$. Events in both the $\ell+{\rm jets}$ and dilepton final states are 
 selected with a multivariate likelihood discriminant that uses both kinematic and $b$-lifetime information to distinguish $t\bar{t}$ signature from background. We select a sample of 786 $\ell$+jets (175 dilepton) events with an expected background contribution of 225$\pm$10 (51$\pm$8) events. Template distributions of cos$\theta^*$ are determined for the $t\bar{t}$ signal using Monte Carlo (MC) simulation of top events, and for the backgrounds using a combination of MC simulation and data. 
To extract $f_0$ and $f_+$, we compute a binned Poisson likelihood $L(f_0, f_+)$ for the data to be consistent with the sum of signal and background templates at any given value for these fractions. The background normalization is constrained to be consistent within uncertainties with the expected value by a Gaussian term in the likelihood. Systematic uncertainties are evaluated in ensemble tests by varying the parameters that can affect the shapes of the cos$\theta^*$ distributions or the relative contribution from signal and background sources. We measure: $f_0 = 0.490\pm0.106~(\rm stat.)\pm0.085~(\rm syst.)$ and $f_+ = 0.110\pm0.059~(\rm stat.)\pm0.052~(\rm syst.)$.
The data indicates fewer longitudinal and more right-handed $W$ bosons than the SM predicts, but is compatible with the SM within the experimental uncertainties. The most precise determination of $f_0$ and $f_+$ has so far been obtained~\cite{cdfhel} by fixing $f_+$ ($f_0$) to the SM expectation and measuring $f_0$ ($f_+$), and yields: $f_0=0.62\pm0.11$ and $f_+=-0.04\pm0.05$.

\noindent {\it \underline {Other Properties}}\\
The measurement of the top quark charge $q$ has been performed using 370 pb$^{-1}$ of D0 collision data,~\cite{topcharge} excluding $|q|=4e/3$ with $\sim$90\% confidence level (C.L.); the result has been recently confirmed by the CDF collaboration. The top quark width has been measured~\cite{topwidth} using 1 fb$^{-1}$ of data collected with the CDF II detector, as $\Gamma_{\rm top}<13.1$~GeV/$c^2$ with 95\% C.L., which is about an order of magnitude larger than the predicted width. Finally, an experimental constraint of the Cabibbo-Kobayashi-Maskawa matrix element $|V_{\rm tb}|>0.89$ with 95\% C.L. has been reported~\cite{vtb} using 0.9 fb$^{-1}$ of data collected with the D0 detector.

\section{\boldmath Conclusions}
A number of properties of the top quark, other than its mass and pair production cross section, have been measured but still suffer from large uncertainties. Indeed, the experimental determination of many of these properties, including for example the top quark production mechanism, the forward-backward asymmetry, the differential cross-section and the helicity of $W$ bosons, are only recently approaching the accuracy needed to test the SM. Nevertheless, all measurements presented in this article indicate that the dynamics of top production and decay at the Tevatron is in agreement with the SM expectation. Moreover, since all results are currently limited by the statistical uncertainties, increased accuracy is expected in the near future by the analysis of all available data sets, and exploiting the additional $\sim$3--4 fb$^{-1}$ of integrated luminosity foreseen at the Tevatron collider by the end of 2010.

\end{document}